\documentclass[12pt,a4paper]{article}
\usepackage{amsmath}
\usepackage{amsfonts}
\usepackage{cite}
\usepackage{graphicx}
\usepackage{epsfig,multicol,bbm}

\begin{document}
\begin{titlepage}

\vfill
\begin{flushright}
ACT-08-10, MIFPA-10-25
\end{flushright}

\vfill

\begin{center}
   \baselineskip=16pt
   {\Large\bf $N=2$ Generalized  Superconformal Quiver Gauge Theory}
   \vskip 2cm
     Dimitri Nanopoulos $^{1,2,3}$, and Dan Xie $^{1}$       \vskip .6cm
      \begin{small}
      $^1$\textit{ George P. and Cynthia W.Mitchell Institute for Fundamental Physics,
Texas A\&M University, College Station, TX 77843.}
        \end{small}\\*[.6cm]
      \begin{small}
      $^2$\textit{Astroparticle physics Group, Houston Advanced Research
Center (HARC), Mitchell Campus, Woodlands, TX 77381, USA.}
        \end{small}\\*[.6cm]
       $^3$\textit{Academy of Athens, Division of Nature Sciences, 28
panepistimiou Avenue, Athens 10679, Greece.}
\end{center}

\vfill
\begin{center}
\textbf{Abstract}\end{center}

\begin{quote}Four dimensional $N=2$ generalized superconformal field theory can be defined by
compactifying six dimensional $(0,2)$ theory on a Riemann surface with regular punctures.
In previous studies, gauge coupling constant space
is identified with the moduli space of punctured Riemann surface $M_{g,n}$. We show
that the weakly coupled gauge group description corresponds to a stable nodal curve, and
the coupling space is actually the Deligne-Mumford compactification $\bar{M}_{g,n}$.
We also give an algorithm to determine the weakly coupled gauge group and matter content in any duality frame.
 \end{quote}

\vfill

\end{titlepage}

\section{Introduction}
Four dimensional  $N=2$ superconformal $A$ type quiver gauge theory (the quiver has the shape
of  $A_n$ type dynkin diagram) is realized as the
six dimensional $(0,2)$ SCFT compactified on a punctured Riemann surface \cite{Argy, Gaiotto1}.
The gauge coupling constants of four dimensional theory
are identified with the complex structure moduli of the punctured Riemann
surface $M_{g,n}$.  The $N=2$ dualities are interpreted geometrically as the conformal mapping
group of the Riemann surface.
It is shown in \cite{moore1, dan} that the BPS equation governing the compactification
is the Hitchin equation \cite{hitchin1,hitchin2} defined on the Riemann surface; At the puncture,
the solution of the Hitchin equation has a regular singularity. The spectral curve of Hitchin's system 
is identified with the Seiberg-Witten fibration \cite{witten4,witten5},  so the IR behavior of the four dimensional
theory can also be determined.

We can engineer a large class of $N=2$ superconformal field theories (SCFT) by
putting a collection of regular singularities on the Riemann surface. We may
call the corresponding gauge theory as the generalized quiver gauge theory.
In general, these theories have no Lagrangian description. However,
It would be interesting to determine the weakly coupled gauge groups and matter content in any duality frame.
Gaiotto \cite{Gaiotto1} argued that the weakly coupled gauge group description
corresponds to the degeneration limit of the Riemann surface, different weakly coupled
descriptions correspond to different degeneration limits of the Riemann surface.
This can be easily applied to the generalized quiver gauge theory.

There is a puzzle why the degeneration limit involves only three punctured sphere (i.e. not two punctured sphere).
Since the gauge coupling constant is identified with the moduli space of the Riemann surface 
and the weakly coupled theory is living on the boundary of the moduli space, we 
show that the weakly coupled description is naturally related to the stable nodal curve which 
is used to compactify the moduli space.
The stable nodal curve is certain kind of singular limit of the punctured Riemann surface and
it represents the boundary point of the moduli space of Riemann surface,
therefore the coupling space is
identified with Deligne-Mumford compactification of moduli space $\bar{M}_{g,n}$ \cite{mirror}.
It is interesting to note that the same compact space plays an important role in 2d
conformal field theory \cite{shenker}. 
We then give an algorithm to calculate the decoupled gauge group and the new puncture
by matching the Coulomb branch moduli. Similarly, by matching the Higgs branch moduli,
we can find the matter content ending on a single gauge group.

This paper is organized as follows: In section II, we review  six dimensional description
of a large class of four dimensional $N=2$ SCFT and focus on counting the Coulomb branch and Higgs branch dimensions; 
In section III, we introduce the definition of stable nodal curve and argue that the weakly coupled description corresponds
to nodal curve; In section IV, we give an algorithm to identify the weakly coupled
gauge group; In section V, we discuss some examples by applying the algorithm developed in
section IV; Finally, we give a discussion about the future
research direction.

\section{Review on counting Coulomb branch and Higgs branch dimensions}
A large class of four dimensional $N=2$ superconformal gauge theories can be engineered as the
six dimensional $(0,2)$ $A_{N-1}$ SCFT compactified on a Riemann surface with or without marked points.
One can argue that the BPS equation for the compactification is Hitchin's equation which 
has the regular singularity at the marked points. 
Let's pick $SU(N)$ gauge group and write $g$ for its lie algebra, $t$ the lie algebra of the maximal torus $T$.
Hitchin's equation is
\begin{eqnarray}
F-\phi\wedge\phi=0\nonumber\\
D\phi=D*\phi=0,
\end{eqnarray}
where $F$ is the curvature of the connection $A_\mu$ of a vector bundle defined on Riemann surface $\Sigma$,
$\phi$ is the one form called Higgs field. The local behavior of conformal invariant solution to Hitchin's equation with regular singularity is \cite{witten2}:
\begin{eqnarray}
A=\alpha d\theta +...\nonumber\\
\phi=\beta {dr\over r}-\gamma d\theta+...
\end{eqnarray}
where $\alpha, \beta, \gamma \in t$ (more precisely, $\alpha$ takes value in maximal torus $T$) and
less singular terms are not written explicitly; $z=re^{i\theta}$ is the local holomorphic coordinate.
The  moduli space of Hitchin's equation with the above behavior around the singularity is denoted as $M_H(\Sigma, \alpha, \beta,\gamma)$. $M_H(\Sigma, \alpha, \beta,\gamma)$ is a hyperkahler manifold and has a family of complex structures parameterized by $CP^1$.  In one distinguished complex structure I,
each point on moduli space represents a Higgs bundle on Riemann surface;
the complex structure modulus is $\beta+i\gamma$, and the kahler modulus is $\alpha$.
In the study of Seiberg-Witten curve of four dimensional theory,
only complex structure of the moduli space matters. since the residue of the Higgs field is $\sigma={1\over 2}(\beta+i\gamma)$ which determines the complex structure of the moduli space, we will focus
on the coefficient of Higgs field. Physically, these parameters are identified with the mass parameters and the complex structure of 
the Riemann surface is identified with the gauge coupling constants, these exhaust all the relevant deformations of the field theory.

The local moduli space of solution is described by the adjoint orbit ${\cal O}_i$ of the complex
lie algebra $sl(N,c)$ (the relation of the adjoint orbit
to moduli space of Nahm's equation can be found in \cite{kro1,kro2}) . The nilpotent orbit is used to describe the massless theory while the semi-simple
orbit is used to describe the mass-deformed theory. The nilpotent orbit is classified by the Young tableaux
$[n_1,n_2,...n_r]$ (for an introduction to nilpotent orbit, see \cite{nil}),
and the mass-deformed theory can be also read from  Young tableaux: there are a
total of $n_1$ mass parameters and the degeneracy of each mass parameter is equal to the number of
boxes on each column . There are only $n_1-1$ independent mass parameters because of the traceless condition.The dimension of the local moduli space is equal to the dimension of the
orbit ${\cal O}_i$:
\begin{equation}
dim({\cal O}_i)=N^2-\sum r_j^2, \label{local}
\end{equation}
where $r_j$ is the height of $j$th column of the Young tableaux.  We can also
read the flavor symmetry from the Young Tableaux, it is
\begin{equation}
S(\prod_{l_h>0}U(l_h)), \label{flavor}
\end{equation}
where $l_h$ is the number of columns with height h, the maximal possible simple subgroup is
$SU(n_1)$. The contribution of this puncture to the Higgs branch moduli is (see the derivation
in appendix A)
\begin{equation}
l_i={1\over 2}(\sum_{k=1}^r n_k^2-N).
\end{equation}
The simple puncture \footnote{Simple puncture has Young Tableaux $[2,1,1,1,..,1]$ and full puncture has Young Tableaux $[N]$.}
 contributes $1$ and the full puncture contribute ${1\over 2}(N^2-N)$. 

The Hitchin's moduli
space on the sphere can be modeled as the quotient
\begin{equation}
({\cal O}_1\times{\cal O}_2...\times{\cal O}_m) / G,
\end{equation}
where $G$ is the complex gauge group, and the total dimension is the sum of the local dimension minus
the dimension of the gauge group:
\begin{equation}
{1\over 2}\sum dim({\cal O}_i)+(g-1)(N^2-1). \label{coulomb}
\end{equation}
Similarly, the
total dimension of the Higgs branch is (In fact, for higher genus case, the number is not
really the dimension of the Higgs branch, since it might be negative; This number is counting
the dimension of the matter minus that of gauge groups.)
\begin{equation}
\sum_il_i+(1-g)(N-1). \label{higgs}
\end{equation}

The above description tells us what is the contribution of a single puncture to
the Coulomb branch, we also want to know what is the total dimension of degree
$i$ operators in Coulomb branch. This can be seen from Seiberg-Witten curve.
The Seiberg-Witten curve is the spectral curve
\begin{equation}
\det(x-\Phi(z))=x^N-\sum_{i=2}^N\phi_i(z)x^{N-i}=0,
\end{equation}
where $\Phi(z)$ is the holomorphic part of the Higgs field and $\phi_i$ is the degree
$i$ meromorphic differential on the Riemann surface parameterized by $z$. For the massless
theory, the pole of order of $\phi_i$ at $j$th puncture is
\begin{equation}
p_i^{(j)}=i-s_i^{(j)}, \label{local1}
\end{equation}
where $s_i^{(j)}$ is the height of $i$th box in the Young tableaux for the $j$ the puncture. The coefficient of
this differential represents dimension $i$ operator parameterizing the Coulomb branch of
quiver gauge theory, the total dimension of this differential is
\begin{equation}
d_i=\sum_{j=1}^np_i^{(j)}-2i+1.
\end{equation}

Various extensions and study of these theories can be found in \cite{SO, web, dan2, N1, yuji2, stony1,stony2, maldacena}.

\section{Nodal curve and weakly coupled gauge theory description}
The gauge coupling constants of four dimensional
$\mathcal{N}=2$ SCFT are identified as the complex structure of a Riemann surface
with punctures, and the S duality group is identified with the modular group. 
The weakly coupled gauge group descriptions correspond to the 
cusp boundary of the moduli space as shown by the original example 
of Argyres and Seiberg. Interestingly, the boundary and the Deligne-Mumford compactification 
of the moduli space is provided by the three punctured sphere and different
cusps corresponds to different three punctured degenerations. Therefore,
the weakly coupled gauge theory description is amazingly related to one of 
the compactification of the moduli space. \footnote{There are some other types  of
compactification like Thurston compactification.}.

Consider a two dimensional topological surface $\Sigma$ with g handles and
n marked points. This manifold can be made into a complex manifold by
defining a complex structure $J$ on it. A complex structure $J$ is a local
linear map on the tangent bundle that satisfies $J^2=-1$ and the integrability
condition. Two complex structures are considered equivalent if they are
related by a diffeomorphism. The moduli space $M_{g,n}$ is the space of all
the inequivalent complex structure on the surface. By Riemann-Roch this is a
space of complex dimension
\begin{equation}
dim M_{g,n}=3g-3+n.
\end{equation}
$M_{g,n}$ is a noncompact complex space with singularities. It arises as
the quotient of a covering space known as Teichmuller space $T_{g,n}$, by a discrete
group, conformal mapping class group $MC_{g,n}$:
\begin{equation}
M_{g,n}={T_{g,n}\over MC_{g,n}}.
\end{equation}
This action typically has fixed points, and the moduli space has orbifold
singularities.

There is another useful way to think about the complex structure on $\Sigma$.
We can think of the point on the moduli space as the conformal class of a metric
$g_{\mu\nu}$. Indeed, a metric defined a complex structure through
\begin{equation}
J_{\mu}^{~\nu}=\sqrt{h}\epsilon_{\mu\lambda}h^{\lambda\nu},
\end{equation}
with $\epsilon_{\mu\nu}$ the Levi-Civita symbol. The definition of the complex
structure does not depend on the local resealing of the metric $g_{\mu\nu}$, so
we can think of the moduli space as the spaces of metric modulo local rescalings
and diffeomorphisms.

The moduli space $M_{g,n}$ is noncompact and has a boundary. The boundary points can be
intuitively represented as degenerate surfaces. The degeneration can be thought
in two ways; the surface can either form a node-or equivalently a long neck- or two
marked points can collide. The process in which two points $x_1$ and $x_2$ collide
if $q=x_1-x_2$ tends to zero can alternatively be described as a process in which a sphere,
that contains $x_1$ and $x_2$ at fixed distance, pinches off the surface by forming a neck of length
$\log q$. So the degeneration limit can be thought of the nodal curve. The boundary points
can be thought of as in the infinity and we would like to compactify this space. The
Deligne-Mumford compactification of $M_{g,n}$ is achieved by adding some points which
represent stable nodal curves.

\begin{figure}
\begin{center}
\includegraphics[width=4.5in]
{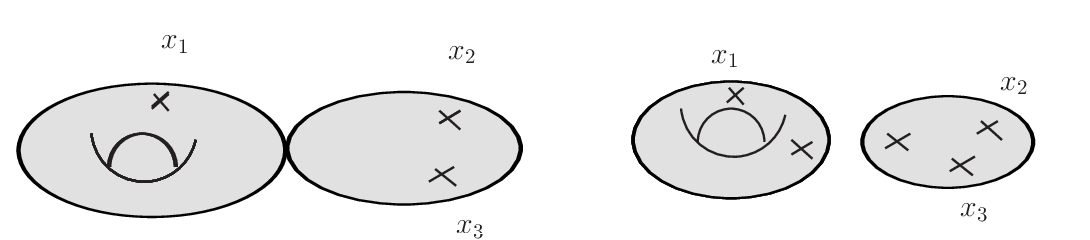}
\end{center}
\caption{Left: A nodal curve. Right: The normalization of a. }
\label{nodal}
\end{figure}

In the following, we will introduce some basic concepts about the nodal curve.
Singular objects play an important role in algebraic geometry. The simplest
singularity a complex curve can have is a node. A nodal point of a curve is a
point that can be described locally by the equation $xy=0$ in $C^2$. An example
is shown in Figure \ref{nodal}a).

We also find the following description of nodal curve very useful. On a surface
with node, the node separates the surface into two components, on the neighborhood
of each node, we can choose local coordinate disks $\{z_i:|z_i|<1\},i=1,2$. The
two disks are glued together at the origin $z_1,z_2=0$ to form the node. We can
open the node by introducing one of complex coordinate $q$ of the moduli space $M_{g,n}$.
Remove the sub-disks $|z_i|<|q|^{1\over2}$ and attach the resulting pair of annuli at
their inner boundaries $|z_i|=|q|^{1\over2}$ by identifying $z_2=q/z_1$. This coordinate
neighborhood on the surface is mapped to a single annulus $|q|^{{1\over2}}<|z|<|q|^{{-1\over2}}$,
by
\begin{eqnarray}
z=q^{1/2}/z_2,~~~if~|q|^{1/2}<|z|\leq 1, \nonumber\\
z=q^{1/2}/z_2,~~~if~1\leq|z|< |q|^{-1/2}.
\end{eqnarray}
As $q=0$, we recover the node. A further transformation $\omega=(2\pi i)^{-1}ln z$ pictures
the opened node as a long tube. Writing $q=e^{2\pi i\tau}$, the length and width is
determined by $\tau$. The node corresponds to a tube of infinite length. In this description,
we see that the moduli is localized on the long tube, and since we identify the moduli
with the gauge coupling constant, we can think that the gauge group is represented by the
long tube.

\begin{figure}
\begin{center}
\includegraphics[width=4.5in]
{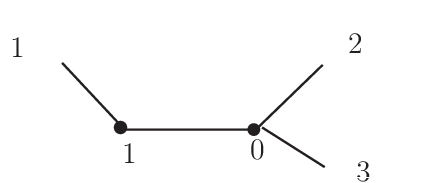}
\end{center}
\caption{ A dual graph for the nodal curve in Figure \ref{nodal}. }
\label{dual}
\end{figure}

We define the normalization of the nodal curve as ungluing its nodes, and add
a marked points to each of the components on which the nodes belong to. See
Figure \ref{nodal}b) for an example. Each component $\Sigma_i$ after the normalization
is an irreducible component of $\Sigma$. There is another convenient way of describing the nodal curve by drawing a
dual graph. The vertices of the dual graph of $\Sigma$ corresponds to components
of $\Sigma$ (and are labeled by their genus), and the edge correspond to node,
we use labeled tails to represent the marked points. An example is shown in Figure \ref{dual}.

A stable nodal curve is a connected nodal curve such that:

(i) Every irreducible component of geometric genus 0 has at least three special points
(including the marked points and the nodal points after the normalization).

ii) Every irreducible component of geometric genus 1 has at least one special
point.

Deligne-Mumford compactification $\bar{M}_{g,n}$ includes the points
corresponding to the stable curve to moduli space $M_{g,n}$.

\begin{figure}
\begin{center}
\includegraphics[width=4.5in]
{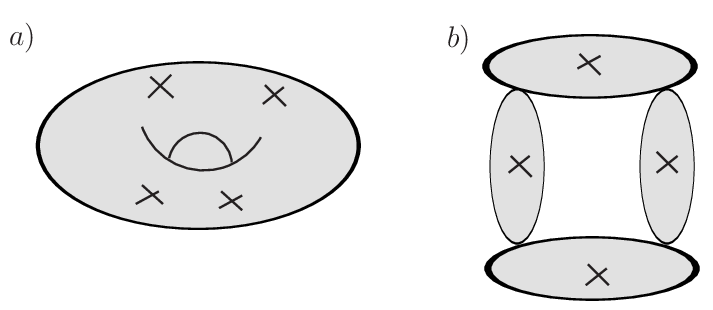}
\end{center}
\caption{ Left: A torus with four marked points. Right: An irreducible nodal curve of a. }
\label{irreducible}
\end{figure}

\begin{figure}
\begin{center}
\includegraphics[width=3.5in]
{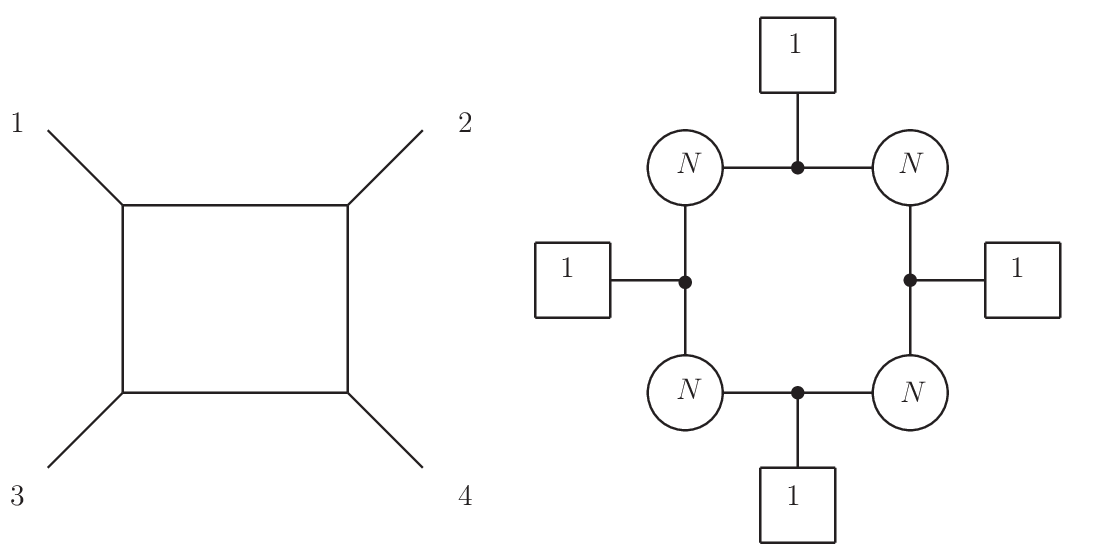}
\end{center}
\caption{ Left: The dual graph for the irreducible nodal curve of Figure \ref{irreducible},
we omit the genus $0$ on each vertex for simplicity, since for irreducible nodal
curve, all components have genus zero. Right: Four dimensional
gauge theory, we put a gauge group on the internal line, external lines
represent the $U(1)$ flavor symmetries. }
\label{dual1}
\end{figure}

Let's define an irreducible nodal curve as a curve whose irreducible components
are all genus 0 curve with three special points. See Figure \ref{irreducible} for an example,
The dual graph for this particular nodal curve is depicted in Figure \ref{dual1}a.

It is time now to connect the nodal curve to the weakly coupled four dimensional
$\mathcal{N}=2$ quiver we are studying in last section. As we reviewed in last section, each
puncture is associated with certain flavor symmetry, and the node or the long neck is
identified with the weakly coupled gauge group, we have the following identification:
A generalized quiver with weakly coupled gauge group associates with the stable nodal
curve and the quiver with all gauge group weakly coupled is the irreducible nodal curve.

It is illuminating to note that $\bar{M}_{g,n}$ also plays an important role
in 2d conformal field theory. It might be natural to expect that the physical quantities calculated 
from the gauge theory side could be related to that of a certain two dimensional conformal field 
theory. A remarkable correspondence is the gauge theory partition function on $S^4$ and the
two dimensional Liouville partition function as conjectured by \cite{Gaiotto2}.  There 
might be some other quantities of the gauge theory which could be related to other two dimensional
CFT.

\section{The gauge group and matter content}
\subsection{Weakly coupled gauge group}

In last section, we established a relation between $\mathcal{N}=2$ weakly coupled SCFT
and the nodal curve and the importance of the three punctured sphere. The remaining task is
to determine what is the weakly coupled gauge groups and matter content in different duality frame.
The matter content is generically a three punctured isolated SCFT and some free matter. These
informations are all encoded in the newly appeared puncture in the normalization of nodal curve.
So we would like to determine what is the newly appeared puncture in any duality frame.

Let's first  define irreducible rank $N$ theory  on  punctured
Riemann sphere from the Seiberg-Witten curve:
\begin{equation}
x^N+\phi_ix^{N-i}=0.
\end{equation}
Here $\phi_idz^i$ is a degree $i$ meromorphic differential defined on  Riemann
sphere, the dimension $d_i$ of this differential is
\begin{equation}
d_i=\sum_j p_i^{(j)}-2i+1,
\end{equation}
here $p_i^{j}$ is the order of pole at the $j$th puncture. $p_i^{(j)}$ can be
read from the Young Tableaux. Consider the degree N
differential, if the number $d_N\leq 0$, then the Seiberg-Witten curve degenerates as
(we consider massless theory here)
\begin{equation}
x(x^{N-1}-\phi_ix^{N-1-i})=0,
\end{equation}
so actually this theory can be realized as a rank $(N-1)$ theory if $d_{N-1}>0$.
We call a theory defined by $A_{N-1}$ compactified on a punctured
Riemann surface irreducible if $d_N>0$ \footnote{
we also include some free theories with $d_N=0$ as irreducible, i.e. the bi-fundamental of $SU(N)$.}.

Now let's discuss the degeneration limit and consider punctured Riemann sphere first. After degeneration,
the original Riemann surface decomposes into two punctured Riemann spheres. From gauge
theory point of view, one of the gauge group is decoupled, and two subquivers are left.
We assume that the decoupled gauge group is a simple gauge group with the
form $SU(k),~k\leq N$ or $USp(2k),~k\leq [{N\over 2}]$. This assumption
will be  confirmed later. The form of the decoupled gauge group is
derived by matching the Coulomb branch moduli of the degeneration limit and the original 
theory.

Let's consider an irreducible rank N theory derived from a Riemann sphere with $n$ punctures, 
and assume one of the gauge group is becoming weakly coupled. Geometrically, this corresponds to
the degeneration of the Riemann sphere. We are left with two punctured spheres $A$ and $B$.
There are two new identical punctures appearing on $A$ and $B$
An important relation between the new puncture and the decoupled gauge group is
that: the decoupled gauge group is a subgroup of the flavor group associated with the new
puncture. Physically, this means the original gauge theory is formed by gauging the subgroup
of the new puncture. Some of the free fields might also decouple in this degeneration limit.

The determination of the new puncture is equivalent to find the 
order of pole of degree $i$ differential at this new puncture.
This is achieved by matching the number of  Coulomb branch moduli with the original quiver. Consider
the degree $i$ moduli, and we assume the original punctures on $A$ contribute to
$\delta_{1i} $ to the Coulomb branch and the original punctures on $B$  has $\delta_{2i}$.
Let's first assume that both components have non-negative degree $i$ moduli, so

\begin{equation}
Case~(1): \delta_{1i}\geq i~~and~~\delta_{2i}\geq i.
\end{equation}

There are two options to consider. First if the decoupled gauge group does not have
a degree $i$ operator, then we have
\begin{equation}
(\delta_{1i}+p_i-2i+1)+(\delta_{2i}+p_i-2i+1)=\delta_{1i}+\delta_{2i}-2i+1,
\end{equation}
where $p_i$ is the contribution from the new puncture to the $i$th degree moduli, this
 gives $2p_i-2i-1=0$ which is inconsistent. On the other hand,
if the decoupled gauge group carry just one degree $i$ moduli (this is the only
choice by our assumption of the decoupled gauge group)
\begin{equation}
(\delta_{1i}+p_i-2i+1)+(\delta_{2i}+p_i-2i+1)+1=\delta_{1i}+\delta_{2i}-2i+1.
\end{equation}
We get $p_i=i-1$. So we conclude that
$p_i=i-1$ with constraint $(1)$. This result is consistent since $A$ and $B$
would also have nonnegative degree $i$ Coulomb branch parameter.

Next let's consider only one set of punctures has degree $i$ moduli, this implies
\begin{equation}
(2): \delta_{1i} \geq i~~ and ~~\delta_{2i}< i; ~or~(3): \delta_{1i}< i~~ and~~ \delta_{2i}\geq i.
\end{equation}

For case $(2)$, $B$  does not have a degree $i$ operator since the maximal contribution of the new appearing puncture
to degree $i$ differential is $p_i=i-1$, and the degree $i$ operator on $B$ is
\begin{equation}
d_i^{(2)}\leq\delta_{2i}+(i-1)-2i+1=\delta_{2i}-i<0.
\end{equation}

There are also two options for the decoupled gauge group. First, if the decoupled gauge group
does not carry a degree $i$ operator, we have
\begin{equation}
(\delta_{1i}+p_i-2i+1)=\delta_{1i}+\delta_{2i}-2i+1.
\end{equation}
This gives $p_i=\delta_{2i}$. 

If the decoupled gauge group has a degree $i$ operator, the equation is
\begin{equation}
(\delta_{1i}+p_i-2i+1)+1=\delta_{1i}+\delta_{2i}-2i+1,
\end{equation}
which gives $p_i=\delta_{2i}-1$. However, we now argue that this is not possible from
gauging the flavor symmetry point of view.  Since $\delta_{2i}< i$, write $\delta_{2i}={i-a}$ with $a\geq 1$.
If $p_i={\delta_{2i}-1}=i-{(a+1)}$, then the $i$th box is at the level $(a+1)\geq 2$ in the Young Tableaux of the new puncture.
Since the decoupled gauge group has a degree $i$ operator, then it
 is at least $SU(i)$ or $USp(i)$ ($USp(i)$ is possible with even $i$ ). However,
the first row $n_1$ of new puncture satisfies $n_1<i$ since the $i$th box is not in the first row, the maximal simple
subgroup of the flavor symmetry is less than $SU(i)$. Therefore, the decoupled gauge group is
large than the flavor group of the new puncture which contradicts our assumption.
The same analysis can be done to case $(3)$.

The situation $\delta_{1i}< i$ and $\delta_{2i}< i$ is excluded since we assume that the original theory
is irreducible.

Combining all the analysis above, we can give a concise formula for $p_i$
\begin{equation}
p_i=min(\delta_{1i},\delta_{2i},i-1).
\end{equation}
and if $min(\delta_{1i}, \delta_{2i})\geq i$, there is a degree $i$ operator
for the decoupled gauge group. 

We next consider the degeneration limit of higher genus theory. Let's study Riemann surface with
genus g and n marked points; there are now three kinds of degeneration: the
genus reduces by one, or two marked points collide
and there are a genus g component and a genus zero component left; Finally
there are a genus $g_1$ and genus $g_2$ components with $g_1$ and $g_2$ are nonzero.

In the first case, there is only a genus $g-1$ surface with $n+2$ marked
points left. Denote the local dimension of the new puncture as $d$, we have
\begin{equation}
{1\over 2}\sum d_i+{1\over 2}(2d)+(g-1-1)(N^2-1)+r={1\over 2}\sum d_i+(g-1)(N^2-1),
\end{equation}
where $r$ is the rank of the decoupled gauge group and $d_i$ is  the dimension of
nilpotent orbit associated with the puncture $i$, (The total dimension of
Hitchin's moduli space on a genus $g$ Riemann surface  is $\sum d_i+2(g-1)(N^2-1)$,
half of this number is the dimension of the Coulomb branch).
Solving the above
equation, we have
\begin{equation}
d=N^2-(r+1).
\end{equation}
the maximal dimension of $d$ is the dimension of  regular
nilpotent orbit and has the dimension $d=N^2-N$, this implies that the minimal
value of $r$ is $N-1$. However, the maximal rank of the decoupled gauge group
is $(N-1)$. We conclude that the decoupled gauge group is $SU(N)$ and the
new puncture is a full puncture. The original theory is assumed to be
irreducible and we can check genus $(g-1)$ theory is also irreducible. 

For the genus $g-1$ theory, we have the dimension of the Coulomb branch
\begin{equation}
d_{g-1}={1\over 2}\sum_id_i+N^2-N+(g-2)(N^2-1)={1\over 2} \sum_i d_i+(g-1)(N^2-1)-(N-1)
\end{equation}
In the case $g>2$, $d_{g-1}>0$ is always true. In the case of $g=1$, since the
minimal dimension of the nilpotent orbit is $2N-2$, we see that $d_{g-1}\geq 0$.
This result shows that the handle of the Riemann surface can only be formed
by a $SU(N)$ group.

The result can also be confirmed by matching Higgs branch moduli using (\ref{higgs}). The
matching condition is
\begin{equation}
\sum_il_i+2l+(1-(g-1))(N-1)-n=\sum_il_i+(1-g)(N-1),
\end{equation}
where $l$ is the contribution of the new puncture and $n$ is the dimension of the
decoupled gauge group, which is $n=(N^2-1)$ in our case. The new puncture is a full puncture
and have $l={1\over 2}(N^2-N)$.

The degeneration limit with genus $g_1$ and $g_2$ parts can be analyzed
similarly. The $g_1$ component has $n_1+1$ marked points and $g_2$ component has
$n_2+1$ marked points, according to our previous analysis, these two theories are
both irreducible. We have the following relation for the coulomb branch dimension
\begin{eqnarray}
\sum k_{1i}+{1\over 2}d+(g_1-1)(N^2-1)+\sum k_{2i}+{1\over 2}d+(g_2-1)(N^2-1) \nonumber\\
=\sum (k_{1i}
+k_{2i})+(g_1-g_2-1)(N^2-1)-r.
\end{eqnarray}
where $r$ is the rank of the decoupled gauge group and $d$ is the dimension of
the nilpotent orbit associated with the puncture as we defined above. Similar analysis
shows that the decoupled gauge group is $SU(N)$ and the new puncture is a full puncture.

The last case with a genus g component and genus zero component is a little bit different.
We know that a genus $g$ component is irreducible, there are nonzero moduli for each
degree. Assume the contribution of two punctures to the moduli of
degree $i$ is $\delta_{1i}$, similar analysis with the degeneration limit of genus
zero case can be done and we have the following conclusion about the order
of poles of the new puncture
\begin{equation}
p_i=min(\delta_{1i}, i-1).
\end{equation}
The decoupled gauge group can be derived by noticing that if $\delta_{1i}\geq i$,
the decoupled gauge group has a degree $i$  operator.

\newpage
\subsection{Matter content}
Let's consider the matter content in each duality frame.  Before doing that, let's discuss 
a little bit more about our general formula for determining the gauge group. First, if 
$p_N=N-1$, then the new puncture $e$ is automatically a full puncture and $p_i=i-1$
for any $i$.  Second, if $p_N<N-1$, then one of the part in the degeneration limit is 
a reducible theory since (assume $\delta_{1N}<(N-1)$ and $p_N=\delta_{1N}$)
\begin{equation}
d^{(1)}_N=\delta_{1N}+p_N-2N+1=2\delta_{1N}-2N+1<0.
\end{equation}

The above two observations tells us that whenever there is a non-full puncture in 
the degeneration limit, there is a reducible theory in one of degeneration part. 

Now let's consider the matter content  of  the first decoupled gauge group 
as depicted in figure. \ref{matter}.  We have already found the gauge group and 
the new appearing puncture $e$.  The matter content is encoded in 
the decoupled three punctured sphere.   However, it is subtle to extract 
the true matter content from the geometry if the decoupled three punctured 
sphere is a reducible theory.  There are three cases which we would like to study in full detail:
\begin{figure}
\begin{center}
\includegraphics[width=5.5in]
{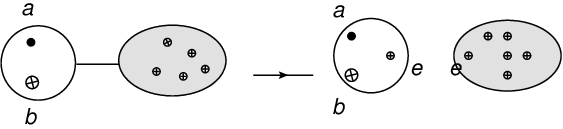}
\end{center}
\caption{ One weakly coupled gauge group }
\label{matter}
\end{figure}

a. The three punctured sphere is an irreducible theory, then the matter content is 
just the isolated SCFT defined by the sphere.

b. The three punctured sphere has some non-zero degree $i$ operators, then 
the matter content is the strongly coupled matter (which is best understood using
the lower rank Hitchin system), plus some free matter.

c. The three punctured sphere has no Coulomb branch operator, and 
there is only free matter.

There is not many new stuff we can say about situation $a$ since the matter is 
just the strongly coupled SCFT. Situation $b$ is a little bit subtle but we could 
extract the strongly coupled matter and its lower rank group  Hitchin representation
using three dimensional mirror symmetry \cite{KS, mirror1, mirror2}, and the free matter part can be 
found by matching the Higgs branch dimensions. Case $c$ is also easy to study
by matching the Higgs branch dimensions.

Let's discuss case $a$ in some further detail to confirm that there is no free matter 
coupled to the gauge group. The match of Higgs branch dimensions gives the 
following equation, we assume the bigger part in the decoupling limit is irreducible and 
the new puncture is then the full puncture, while the decoupled gauge group is $SU(N)$
\begin{equation}
l_a+l_b+l_{\Delta}+(N-1)=(l_a+l_b+l_e+N-1)+(l_{\Delta}+l_e+N-1)-(N^2-1).
\end{equation}
The above equation is valid if 
\begin{equation}
2l_e=N^2-N,
\end{equation}
which is true for the full puncture. The other cases can be worked out quite easily 
as many examples in next section will tell.

\newpage
\section{Examples}
We apply our formula to calculate the decoupled gauge group and matter content for some examples in this section.
We always start with an irreducible theory.

\textbf{Example 1}: The six dimensional description of a linear quiver of $A_N$
type involves two generic punctures, and several
basic punctures. We decide what happens when a simple puncture is colliding with a
generic puncture with rows $n_1\geq n_2...\geq n_k$, and the height of
the first column is $s_1$.  The other part in the degeneration has to be an irreducible theory and so 
the new appearing puncture is determined by two colliding puncture and we 
have the formula
\begin{equation}
 p_i=\min(\delta_{1i}, i-1).
 \end{equation}
 
The data is listed in  Table \ref{tb1}. In the last line of Table \ref{tb1}, we indicate whether decoupled gauge group has a
degree $i$  operator. So the decoupled gauge group is a $SU(n_1)$ group.  The new puncture has the feature that  the
first row and second row are combined and other rows are unchanged.  
\begin{table}
\begin{center}
  \begin{tabular}{ |l | c |c| c|c|c|c|c|r | }
    \hline
    i & 2 & 3& 4 &... & $n_1$ &$n_1+1$&... &N \\ \hline
 $p_{1i}$ & 1 & 1&1&...&1&1&...&1 \\ \hline
    $p_{2i}$&1&2&3&... &$n_1-1$&$n_1-1$&...&$N-s_1$ \\ \hline
    $\delta_{1i}$&2&3&4&...&$n_1$&$n_1$&...&$N-s_1+1$ \\ \hline
    $p_i$&1&2&3&...&$n_1-1$&$n_1$&...&$N-(s_1-1)$ \\ \hline
         &1&1&1&...&1&0&...&0\\
    \hline
  \end{tabular}
  \end{center}
  \caption{The data needed for colliding a generic puncture and a simple puncture. }
  \label{tb1}
\end{table}

Now let's determine the matter content, since the decoupled three punctured sphere has no
Coulomb branch dimension, there are free matter coupled on this decoupled group, let's 
assume the decoupled matter has Higgs branch dimension $x$, then we have
\begin{equation}
\sum_{i=2}^n l_i+l_2+1+(N-1)= \sum_{i=2}^n l_i+l_2+n_1n_2-(n_1^2-1)+(N-1)+x,
\end{equation}

Here $l_2$ is the contribution of the generic puncture, 1 is the contribution of the simple
puncture; we have used the fact that the new puncture has the contribution to Higgs branch
$(l_2+n_1n_2)$, where
$x$ is the contribution from the fundamental fields. Calculate it, we get $x=n_1(n_1-n_2)$, so
there is $n_1-n_2$ fundamentals on $SU(n_1)$.

\textbf{Example 2}: Let's consider collision of two identical punctures which have two
columns with equal height N. We list the analysis in the Table \ref{tb2}:
\begin{table}
\begin{center}
  \begin{tabular}{ |l | c |c| c|c|c|c|c|r | }
    \hline
    i & 2 & 3& 4 &... & $2k$ &$2k+1$&... &2N \\ \hline
 $p_{1i}$ & 1 & 1&2&...&k&k&...&N \\ \hline
    $p_{2i}$&1&1&2&... &$k$&$k$&...&$N$ \\ \hline
    $\delta_{1i}$&2&2&4&...&$2k$&$2k$&...&$2N$ \\ \hline
    $p_i$&1&2&3&...&$2k-1$&$2k$&...&$2N-1$ \\ \hline
         &1&0&1&...&1&0&...&1\\
    \hline
  \end{tabular}
  \end{center}
  \caption{The data needed for collision of two identical punctures with equal height N.}
  \label{tb2}
  \end{table}
From the Table \ref{tb2}, we can see that the new puncture is a full puncture and the
decoupled gauge group has only even rank operator, the natural decoupled gauge group is
$USp(2N)$, one may wonder why USp gauge group appears when we compactify a $A_{2N-1}$
theory on a Riemann surface, this can be done by including a outer automorphism of
the gauge group $SU(2N)$ in the compactification, see \cite{vafa, witten7
}. 
The matter content is a free matter and we can find it using Higgs branch matching:
\begin{equation}
\sum_{i=2}^nl_i+2N+(2N-1)=\sum_{i=2}^nl_i+(2N^2-N)+(2N-1)-2N^2-N+x,
\end{equation}
We have $x=4N$, so we have 2 fundamentals on USp node. 

\textbf{Example 3}:
We also confirm another example which is studied in \cite{dan2}. One puncture has partition
$[2,2...2]$, the other puncture has partition $[3,2....2,1]$, the data is assembled
in Table \ref{tb3}:
\begin{table}
\begin{center}
  \begin{tabular}{ |l | c |c| c|c|c|c|c|r | }
    \hline
    i & 2 & 3& 4 &... & $2k$ &$2k+1$&... &2N \\ \hline
 $p_{1i}$ & 1 & 1&2&...&k&k&...&N \\ \hline
    $p_{2i}$&1&2&2&... &$k$&$k+1$&...&$N$ \\ \hline
    $\delta_{1i}$&2&3&4&...&$2k$&$2k+1$&...&$2N$ \\ \hline
    $p_i$&1&2&3&...&$2k-1$&$2k$&...&$2N-1$ \\ \hline
         &1&1&1&...&1&1&...&1\\
    \hline
  \end{tabular}
  \end{center}
  \caption{The data needed for colliding two punctures appearing in $SU(N)$ theory with antisymmetric matter. }
  \label{tb3}
  \end{table}
From the Table \ref{tb3}, we conclude that the new puncture is a full puncture and the decoupled
gauge group is a $SU(2N)$ gauge group. The three punctured sphere does not
contribute to higgs branch, we have the equation
\begin{equation}
\sum_{i=2}^nl_i+N+(N+1)+(2N-1)=\sum_{i=2}^nl_i+(2N^2-N)+(2N-1)-((2N)^2-1)+x,
\end{equation}
we find $x=2N^2+N$, this sounds weird, since no single conventional matter on $SU(N)$ can
give this number for Higgs branch. However, let's split $2N^2+3N=(2N^2-N)+4N$, that's an
antisymmetric matter and two fundamentals' contribution, it splits in this way so that
the $SU(N)$ gauge group is conformal.

\begin{figure}
\begin{center}
\includegraphics[width=4.0in]
{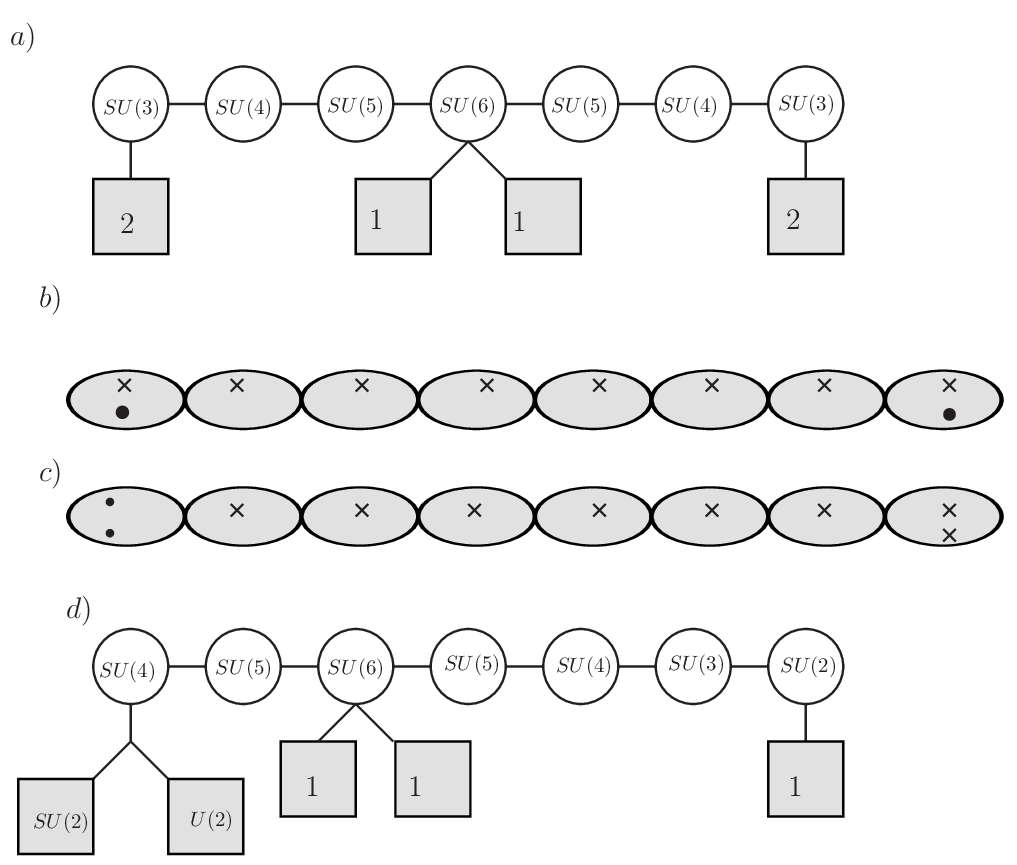}
\end{center}
\caption{a): A linear quiver. b): The six dimensional construction corresponding to quiver in (a), the
cross denotes the simple puncture and the black dot denotes the puncture with partition
$[3,1,1,1]$. c): A different weakly coupled gauge group description, we collide two generic punctures.
d): A generalized quiver corresponding to (c).}
\label{degeneration}
\end{figure}

\textbf{Example 4}: Finally, let's consider an example of collision of two generic punctures with partitions $[3,1,1,1]$.
The linear quiver gauge theory with these two punctures is depicted in Figure \ref{degeneration}a. The
six dimensional construction is depicted in Figure \ref{degeneration}b. We study another weakly coupled
 theory corresponding to collide two generic puncture represented by black dot, and
 nodal curve and the generalized quiver is depicted in Figure \ref{degeneration}c and \ref{degeneration}d.
 The analysis is listed in Table \ref{tb4}. The new appearing puncture has the partition $[5,1]$, the decoupled
gauge group is $SU(4)$. The decoupled three punctured sphere is reducible
but it carries a degree $3$ moduli, so there are strongly coupled SCFT and the free matter both
coupled to the gauge group. One can use the 3d mirror method to find that the isolated SCFT
is actually $E_6$ theory which is realized by $A_2$ on a sphere with three punctures. Here the 
$SU(4)$ subgroup is gauged.
 
\begin{table}
\begin{center}
  \begin{tabular}{ |l | c |c|c|c|r | }
    \hline
    i & 2 & 3& 4 & 5&6 \\ \hline
 $p_{1i}$ & 1 & 2&2&2&2 \\ \hline
    $p_{2i}$&1&2&2&2&2 \\ \hline
    $\delta_{1i}$&2&4&4&4&4 \\ \hline
    $p_i$&1&2&3&4&4\\ \hline
         &1&1&1&0&0\\
    \hline
  \end{tabular}
  \end{center}
  \caption{The data needed for colliding two generic punctures.}
  \label{tb4}
  \end{table}

Now let's determine the free matter part. The total dimension of Higgs
branch of quiver depicted in Figure \ref{degeneration}a) is 19 using our formula (\ref{higgs})(or just from quiver). The $E_6$ theory 
has Higgs branch dimension 11. We have the following Higgs branch matching condition:
\begin{equation}
23+11+x-15=19,
\end{equation}
where $23$ is from the left quiver and $x$ is the contribution from the free matter
fields, $15$ is the dimension of the decoupled gauge group.
We get $x=0$ which means there is no extra fundamentals on $SU(4)$ node.

\section{Conclusion}
In this paper, we studied four dimensional $N=2$ generalized superconformal quiver gauge theory which
is derived from six dimensional $(0,2)$ SCFT compactified on a Riemann surface with
some punctures. The weakly coupled gauge theory description is living at the boundary
of the moduli space of the Riemann surface and therefore is naturally related 
to the compactification of the moduli space. We show that the Deligne-Mumford
compactification is the right one for the gauge theory description which provide 
a confirmation of earlier realization that the weakly coupled gauge group 
description corresponds to the degeneration of the Riemann surface into three 
punctured sphere. 

We give an explicit formula to determine the weakly coupled gauge group in any duality frame, 
and an explicit algorithm for determining the matter content is also presented. Therefore we 
have explicit gauge group and matter content information in any duality frame. One
could easily confirm our general analysis with the examples presented in \cite{distler}.
We believe that the method we present here can also be applied to four
dimensional $N=2$ $A$ type quiver with $USp-SO$ group \cite{SO, SO1, SO2}.

\begin{flushleft}
\textbf{Acknowledgments}
\end{flushleft}
It is a pleasure to thank Yu-Chieh Chung for helpful discussions.
This research was supported in part by the Mitchell-Heep chair in
High Energy Physics (CMC), by the Cambridge-Mitchell Collaboration
in Theoretic Cosmology, and by the DOE grant DEFG03-95-Er-40917.

\appendix
\section{The Contribution to Higgs Branch of One Generic Puncture}

The Higgs branch of the generalized quiver is a little bit difficult. For simplicity,
we first study the Riemann sphere two generic punctures and $s_1+s_2$ simple punctures, we have
the explicit Lagrangian description in one dual frame. Let's first
consider the contribution of the simple puncture. Adding one simple puncture, in the
corresponding quiver, we add a $SU(N)$ node to the original quiver, we need to add a
bifundamental matter between two $SU(N)$ nodes, so the net contribution to the
Higgs branch is $N^2-(N^2-1)=1$.

Now consider a generic
puncture labeled by $[n_1, n_2,...n_s]$ and the quiver tail for it is $SU(k_1)-SU(k_2)-...-SU(N)$, where $k_j=\sum_{i=1}^jn_j$.
For present purpose,
we artificially split the $SU(N)$ gauge group into two equal parts, the "rank" of the gauge group in this
tail is ${1\over 2}(N^2-1)$. There are  $N-k_{s-1}$ fundamentals on the $SU(N)$ gauge group.
The contribution from bifundamental matter of this quiver tail to Higgs branch is
\begin{equation}
\sum_{j=1}^{s-1}k_jk_{j+1}.
\end{equation}
The contribution from the fundamental matter is
\begin{equation}
\sum_{j=1}^{s-1}(2k_j-k_{j-1}-k_{j+1})k_j+N(N-k_{s-1}),
\end{equation}
where $k_j$ is the rank of $j$th gauge group and we set $k_0=0, k_m=N$, $k_j$ is
related to the Young tableaux as $k_j=n_1+...n_j$. The Higgs branch of this tail
is (we assume the gauge group is completely higgsed in Higgs branch)
\begin{equation}
\sum_{j=1}^{s-1}k_jk_{j+1}+\sum_{j=1}^{s-1}(2k_j-k_{j-1}-k_{j+1})k_j+N(N-k_{s-1})-\sum_{j=1}^{s-1}(k_j^2-1)-{1\over 2}(N^2-1).
\end{equation}
After some calculation, we have
\begin{equation}
\sum_{j=1}^{s}k_j(k_j-k_{j-1})+s-1-{1\over 2}(N^2-1)={1\over2}(\sum_{i=1}^{s}n_i^2-N)+s +{N-1\over2}.
\end{equation}
The first term in above form is thought to be the contribution to Higgs branch due to this generic puncture.
Since for this tail, we need $s$ simple punctures, and the contribution of the simple puncture to Higgs branch is 1, so
the second term is thought of the contribution form simple punctures. The last term is thought to be the global contribution
just like $-(N^2-1)$ term for the Coulomb branch. Finally, adding two tails, there are a (N-1) extra terms, which we
think as the total global contribution on the sphere . The total dimension for the Higgs branch is then
\begin{equation}
\sum_i l_i +N-1,
\end{equation}
where
\begin{equation}
l_i={1\over2}(\sum_{j=1}^{s}n_j^2-N).
\end{equation}

For other generalized quiver, since we do not have a Lagrangian description, we do not know how to count the dimension
of Higgs branch. However,
based on the analysis for the linear quiver, we conjecture that for each puncture, the contribution to the Higgs branch is
\begin{equation}
l_k={1\over2}(\sum_{i=1}^sn_i^2-N),
\end{equation}
and there is a $N-1$ global contribution to Higgs branch. The total dimension of Higgs branch is
\begin{equation}
\sum_il_i+(N-1).
\end{equation}
This formula can also be seen from the three dimensional mirror theory \cite{mirror1}.

The generalization to higher loop case is straightforward
\begin{equation}
\sum_il_i+(1-g)(N-1).
\end{equation}
In fact, for the higher loop case, we really count the difference between the dimension
of the matter and the dimension of the gauge group, and in some cases, the theory is
not completely higgsed, there are some unbroken $U(1)$ gauge symmetry.


\begin{thebibliography}{999}
\bibitem{Argy}    P.C.Argyres and N.Seiberg, "S-Duality in $N=2$
Supersymmeric Gauge Theories," JHEP 0712 (2007) 088
[arXiv:0711.0054][hep-th].
\bibitem{Gaiotto1} D. Gaiotto, N=2 dualities, arXiv:0904.2715.
\bibitem{moore1} D.Gaiotto, G.W.Moore, A.Neitzke, Wall-crossing, Hitchin Systems, and the WKB Approximation,
[arXiv:0907.3987].
\bibitem{dan}D.V.Nanopoulos, D.Xie, Hitchin Equation, Singularity, and N=2 Superconformal Field Theories,
JHEP 1003:043,2010 [arXiv:0911.1990].
\bibitem{hitchin1}N.Hitchin, The self-duality equation on a riemann surface, Proc.Lomdon Math.Soc. (3) 55
(1987)59-126.
\bibitem{hitchin2}N.Hitchin, Stable bundles and integrable systems, Duke Math. J. Volume 54, Number 1 (1987), 91-114.
\bibitem{witten4} N.Seiberg and E.Witten, "Monopoles, Duality and
Chiral Symmetry breaking in $N=2$ Supersymmetric QCD," Nucl.Phys. B
431 (1994) 484 [arXiv:hep-th/9408099].
\bibitem{witten5} N.Seiberg and E.Witten, Monopoles, Duality and Chiral Symmetry Breaking in N=2 Supersymmetric QCD,
Nucl.Phys.B431:484-550,1994, [arXiv:hep-th/9408099].



\bibitem{mirror}K. Hori, S. Katz, A. Klemm, R. Pandharipande, R. P. Thomas, C. Vafa, R. Vakil, and
E. Zaslow, Mirror symmetry, vol. 2 of Clay Mathematics Monographs. American
Mathematical Society, Providence, RI, 2003.
\bibitem{shenker}D.Friedan, S.Shenker, The Analytic Geometry of Two-Dimensional Conformal Field Theory.
Nucl.Phys.B281:509,1987.


\bibitem{witten2} S.Gukov and E.Witten, Gauge Theory, Ramification
and the Geometric Langlands Program, [arXiv:hep-th/0612073].


\bibitem{kro1} P.Kronheimer, Instantons and the geometry of the nilpotent variety, J.Diff.Geom.32 (1990) 473-490.
\bibitem{kro2}P.Kronheimer, A hyper-kahlerian structure on coadjoint orbits of a semisimple complex group,
J.London Math.Soc.42(1990)193-208.
\bibitem{nil}D.Collingwood, W.McGovern, Nilpotent orbits in semisimple lie algebra, VanNostrand Reinhold Math.Series,
New York, 1993.























 \bibitem{maldacena}D. Gaiotto, J. Maldacena, The gravity duals of N=2 superconformal field theories,
 [arXiv:0904.4466].
\bibitem{SO} Y. Tachikawa, Six-dimensional DN theory and four-dimensional SO-USp quivers, JHEP 07
(2009) 067, [arXiv:0905.4074].
\bibitem{web} F. Benini, S. Benvenuti, and Y. Tachikawa, Webs of five-branes and N=2 superconformal field
theories, JHEP 09 (2009) 052, [arXiv:0906.0359].
\bibitem{N1}K. Maruyoshi, M. Taki, S. Terashima, and F. Yagi, New Seiberg Dualities from N=2 Dualities,
JHEP 09 (2009) 086, [arXiv:0907.2625].
\bibitem{dan2}D. Nanopoulos and D. Xie, N=2 SU Quiver with USP Ends or SU Ends with Antisymmetric
Matter, JHEP 08 (2009) 108, [arXiv:0907.1651].
\bibitem{yuji2}F. Benini, Y. Tachikawa, and B. Wecht, Sicilian gauge theories and N=1 dualities,
arXiv:0909.1327.
\bibitem{stony1}A.Gadde, E.Pomoni, L.Rastelli, S.S.Razamat, S-duality and 2d Topological QFT,
	JHEP 1003:032,2010, [arXiv:0910.2225[hep-th]].
\bibitem{stony2}A.Gadde, L.Rastelli, S.S.Razamat, W.Yan, The Superconformal Index of the $E_6$ SCFT,
[arXiv:1003.4244[hep-th]].

\bibitem{Gaiotto2} L. F. Alday, D. Gaiotto, and Y. Tachikawa, Liouville Correlation Functions from
Four-dimensional Gauge Theories, Lett.Math.Phys.91:167-197,2010 [arXiv:0906.3219].

\bibitem{KS} K. Intriligator, N. Seiberg, Mirror Symmetry in Three Dimensional Gauge Theories, Phys.Lett.B387:513-519,1996, arXiv:hep-th/9607207.

\bibitem{mirror1}{F. Benini, Y. Tachikawa, D. Xie, Mirrors of 3d Sicilian theories, JHEP 1009:063,2010, arXiv:1007.0992.}

\bibitem{mirror2}{D. Nanopoulos, D. Xie, More Three Dimensional Mirror Pairs, JHEP 1105:071,2011, arXiv:1011.1911.}


\bibitem{vafa}C.Vafa, Geometric Origin of Montonen-Olive Duality, Adv.Theor.Math.Phys.1:158-166,1998,
arXiv:hep-th/9707131.
\bibitem{witten7}E.Witten, New ``Gauge'' Theories In Six Dimensions, JHEP 9801:001,1998; Adv.Theor.Math.Phys.2:61-90,1998, arXiv:hep-th/9710065.




\bibitem{distler}O. Chacaltana, J. Distler, Tinkertoys for Gaiotto Duality, JHEP 1011:099,2010, arXiv:1008.5203.


\bibitem{SO1}K.Landsteiner, E.Lopez, D.A.Lowe, N=2 Supersymmetric Gauge Theories, Branes and Orientifolds,
Nucl.Phys. B507 (1997) 197-226,  arXiv:hep-th/9705199.
\bibitem{SO2}A. Brandhuber, J. Sonnenschein, S. Theisen, S. Yankielowicz, M Theory And Seiberg-Witten Curves: Orthogonal and Symplectic Groups,
Nucl.Phys. B504 (1997) 175-188, arXiv:hep-th/9705232.





\end{thebibliography}
\end{document}